\documentclass[preprint]{emulateapj}

\let\url\relax
\usepackage
[dvips]
{hyperref}
\usepackage[all]{hypcap}
\makeatletter
\renewcommand{\fps@figure}{tp}
\makeatother

\setkeys{Gin}{width=\linewidth,height=0.9\textheight,keepaspectratio=true}

\newcommand\thor[2]{$\theta^{#1}$Ori~#2}
\newcommand{\thC}{\thor{1}{C}}
\newcommand{\thA}{\thor{2}{A}}
\newcommand{\kms}{\ensuremath{\mathrm{km\ s}^{-1}}}
\newcommand{\pcc}{\ensuremath{\mathrm{cm}^{-3}}}

\newcounter{ionstage}
\renewcommand{\ion}[2]{\setcounter{ionstage}{#2}%
  \ensuremath{\mathrm{#1\,\scriptstyle\Roman{ionstage}}}}
\newcommand\OIlam{[\ion{O}{1}]\,6300\,\AA\@}
\newcommand\SIIlam{[\ion{S}{2}]\,6731\,\AA\@}
\newcommand\SIIlamboth{[\ion{S}{2}]\,6716,6731\,\AA\@}
\newcommand\SIIIlam{[\ion{S}{3}]\,6312\,\AA\@}

\newcommand\OI{[\ion{O}{1}]}
\newcommand\SII{[\ion{S}{2}]}
\newcommand\SIII{[\ion{S}{3}]}
\newcommand\OIII{[\ion{O}{3}]}
\newcommand\NII{[\ion{N}{2}]}

\newcommand\HH[1]{HH~#1}
\newcommand\Vomc{\ensuremath{V_\mathrm{OMC}}}
\newcommand\Vt{\ensuremath{V_\mathrm{T}}}
\newcommand\Vs{\ensuremath{V_\mathrm{S}}}
\newcommand\htwo{H$_2$}
\newcommand\Halpha{H$\alpha$}

\newcommand\OW[2]{#1--#2}
\newcommand\xy[2]{($#1$, $#2$)}

\newcommand\Vhel{\ensuremath{V_\odot}}

\shorttitle{Large Scale Flows from Orion-South}
\shortauthors{Henney et al.}

\begin{document}
\makeatletter\global\NAT@numbersfalse\makeatother

\title{Large Scale Flows from Orion-South\footnotemark[1,2,3]}

\footnotetext[1]{Based on observations obtained at the
  \protect\facility{Observatorio Astron\'omico Nacional, San Pedro
    M\'artir, Baja California, Mexico}, which is operated by the
  Universidad Nacional Aut\'onoma de M\'exico.}

\footnotetext[2]{Based on observations obtained at the
  \protect\facility{Kitt Peak National Observatory}, which is operated
  by the Association of Universities for Research in Astronomy, Inc.,
  under a Cooperative Agreement with the National Science Foundation.}

\footnotetext[3]{Based on observations with the NASA/ESA
  \protect\facility{Hubble Space Telescope}, obtained at the Space
  Telescope Science Institute, which is operated by the Association of
  Universities for Research in Astronomy, Inc., under NASA Contract
  No.~NAS 5-26555.}

\author{
  W. J. Henney,\altaffilmark{4}
  C. R. O'Dell,\altaffilmark{5}
  Luis A. Zapata,\altaffilmark{4} 
  Ma.\@ T. Garc\'{\i}a-D\'{\i}az,\altaffilmark{4,6}  
  Luis F. Rodr\'{\i}guez,\altaffilmark{4}\\
  and Massimo Robberto\altaffilmark{7}
} 

\email{w.henney@astrosmo.unam.mx}

\altaffiltext{4}{Centro de Radioastronom\'{\i}a y Astrof\'{\i}sica,
  Universidad Nacional Aut\'onoma de M\'exico, Apartado Postal 3-72,
  58090 Morelia, Michaoac\'an, M\'exico}
\altaffiltext{5}{Department of Physics and Astronomy,
  Vanderbilt University, Box 1807-B, Nashville, TN 37235}
\altaffiltext{6}{Instituto de Astronom\'\i{}a, Universidad Nacional
  Aut\'onoma de M\'exico, Apartado Postal 877, 22800 Ensenada, Baja
  California, M\'exico}
\altaffiltext{7}{Space Telescope Science Institute,
  3700 San Martin Drive, Baltimore, MD 21218}

\begin{abstract}
  Multiple optical outflows are known to exist in the vicinity of the
  active star formation region called Orion-South (Orion-S).  We have
  mapped the velocity of low ionization features in the brightest part
  of the Orion Nebula, including Orion-S, and imaged the entire nebula
  with the Hubble Space Telescope.  These new data, combined with
  recent high resolution radio maps of outflows from the Orion-S
  region, allow us to trace the origin of the optical outflows. It is
  confirmed that HH~625 arises from the blueshifted lobe of the CO
  outflow from \OW{136}{359} in Orion-S while it is likely that HH~507
  arises from the blueshifted lobe of the SiO outflow from the nearby
  source \OW{135}{356}. It is likely that redshifted lobes are
  deflected within the photon dominated region behind the optical
  nebula. This leads to a possible identification of a new large shock
  to the southwest from Orion-S as being driven by the redshifted CO
  outflow arising from \OW{137}{408}.  The distant object HH~400 is
  seen to have two even further components and these all are probably
  linked to either HH~203, HH~204, or HH~528. Distant shocks on the
  west side of the nebula may be related to HH~269. The sources of
  multiple bright blueshifted Herbig-Haro objects (HH~202, HH~203,
  HH~204, HH~269, HH~528) remain unidentified, in spite of earlier
  claimed identifications. Some of this lack of identification may
  arise from the fact that deflection in radial velocity can also
  produce a change in direction in the plane of the sky. The best way
  to resolve this open question is through improved tangential
  velocities of low ionization features arising where the outflows
  first break out into the ionized nebula.
\end{abstract}


\section{Introduction}
As the nearest region of star formation that includes young massive
stars, the Orion Nebula and its eponymous cluster attract considerable
attention from students of star formation. The cluster is displaced
towards the observer from the remainder of the host molecular star and
the visible nebula is a relatively thin, irregular, concave layer of
ionized gas on the observer's side of the molecular cloud. Numerous
optical shocks and jets \citep{ode01} reveal the presence of
collimated and broad outflows from those stars that lie in the
photoionized portion of this region; but one must rely on infrared,
x-ray, and radio observations to tell what is happening among the
optically obscured cluster stars lying beyond the H~II region, which
constitute roughly half of the total cluster membership
\citep{1997AJ....113.1733H}. 
%% BOB: I have rewritten this part slightly since the ONC is generally
%% taken to be the large scale grouping that includes all the stars
%% associated with the OMC-1 cloud, with a size of several
%% parsecs. The BN-KL sources would certainly then be part of the ONC.
%%
%% Original sentence from Bob:
% Not all of these stars are members of the Orion Nebula
% Cluster, rather, they are members of secondary star formation
% groups.
In the case of the highly embedded (about 0.2 pc behind the nebula)
BN-KL region, one only sees a few optical features resulting from
stellar outflow at the tips of the Allen and Burton fingers
(\citeyear{all93}). However, in the region known as Orion-South
(Orion-S, also known as Orion Molecular Cloud~1 South or OMC-1S),
lying 1\arcmin\ to the southwest from the dominant ionizing star \thC,
the embedded sources lie near the surface of the nebula and we see
numerous Herbig-Haro (HH) objects there. This is the third region of
star formation in the vicinity of the Orion nebula and is in a local
bump in the surface of the nebula \citep{wen95}, where the expanding
H~II region has been slowed because of the higher density of gas
there. There have been numerous recent papers that try to relate the
moving optical features to the obscured sources of the outflows, but
we reassess the situation here because of the recent availability
\citep{zap05,zap06} of high spatial resolution maps of CO and SiO
outflows from Orion-S. We present new results obtained
spectroscopically and with imaging in \S~2, summarize what we know
about the optical features in \S~3, present a combined picture of the
optical features in \S~4, and discuss their relation to the CO and SiO
outflows in \S~5.  Throughout this paper we will adopt a distance of
460 pc, after \citet{bom00}, henceforth BOM.\defcitealias{bom00}{BOM}

\vspace*{\baselineskip}
\section{Observations}
In this section we present new observations of the Orion Nebula. The
first results are from a program of mapping the radial velocity across
the Huygens region of the nebula, the second are results from imaging
an even wider region with the Hubble Space Telescope.

\subsection{HH 528}

\begin{figure}
  \centering
  \includegraphics{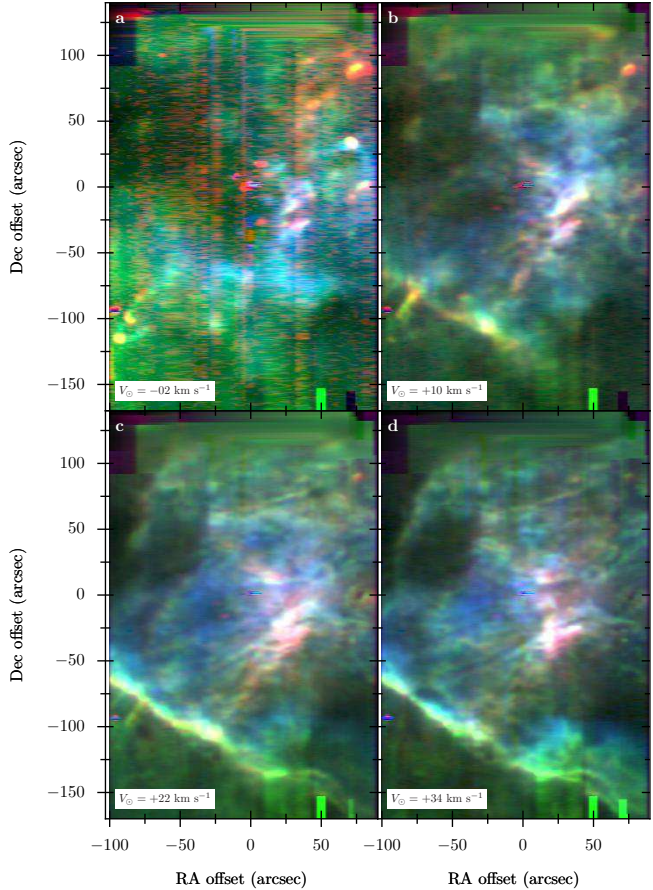} % isomamulti
  \caption[]{Multiline isovelocity maps of the Orion Nebula in which the
    emission in \OI{}, \SII{}, and \SIII{} is shown in red, green, and
    blue, respectively. Each panel shows emission in a 12~\kms{} wide
    channel centered on the heliocentric velocity indicated in the
    lower-left corners. The intensity scalings vary between the
    different channels and different lines. Position offsets are
    measured with respect to \thC{}. }
  \label{fig:isomaps}
\end{figure}

We have constructed isovelocity channel maps of the core of the Orion
Nebula, based on echelle longslit observations in the optical emission
lines of \SIIlamboth{}, \OIlam{}, and \SIIIlam{}. Full details of the
observations and data reduction process are given in \citet[henceforth
GH07]{gar06}\defcitealias{gar06}{GH07}.  The data were obtained
primarily with the Manchester Echelle Spectrometer \citep{mea03} on
the 2.1~m telescope of the Observatorio Astron\'omico Nacional at San
Pedro Martir, B.C., Mexico, while part of the \SII{} dataset was
obtained with the 4~m telescope at Kitt Peak National Observatory
\cite[henceforth DOH04]{doi04}\defcitealias{doi04}{DOH04}.
Figure~\ref{fig:isomaps} shows images of selected channel maps, with
\OIlam{} shown in red, \SIIlam{} shown in green, and \SIIIlam{} shown
in blue. All velocities in this section are given in the heliocentric
frame, and can be transformed to the Local Standard of Rest by
subtracting 18.1~\kms. All positions in this section are given in
arcseconds, \xy{x}{y}, with respect to \thC{}, the brightest of the
Trapezium stars and the principal source of ionizing photons in the
nebula. Some artifacts are present in the data, principally due to
variations in sky brightness, which manifest themselves as vertical
stripes in the images. These are most noticeable where the signal is
intrinsically weak, such as in the blueshifted \OI{} emission (shown
as red in Fig.~\ref{fig:isomaps}\textit{a}) and are further discussed
in \citetalias[Section~2]{gar06}.

The difference in the emission distribution between the three lines is
due primarily to changes in ionization: the \SIII{} line is emitted
only by fully ionized gas, whereas the \OI{} line comes from a thin
layer of partially ionized gas at the ionization front or from shocks
in neutral gas. The \SII{} line comes from both the ionization front
(IF) and from the fully ionized gas that is close to it. In addition,
collisional deexcitation causes the \SII{} line to saturate in the
highest density regions. Since the electron density shows a general
decline with distance from the center of the nebula, the outer parts
of the map have a green tinge, indicative of relatively strong \SII{}
emission, while the inner parts vary from pink to blue, depending on
whether low- or high-ionization emission is dominant.

The emission in the most blueshifted channel that we show, centered on
$\Vhel = -2~\kms$, is dominated by a broad band of high-ionization gas
known as the Big Arc \citep{ode97a}, which crosses the south of the
nebula from east to west. One also sees emission from the low-velocity
wings of the highly blue-shifted HH objects \citep{ode97b}:
\HH{203/204} around \xy{-90}{-100}; \HH{202} around \xy{70}{30}; and
the low-ionization \HH{201} at \xy{75}{90}. As one passes to more
redshifted channels, these features fade away to be replaced by the
systemic nebular emission, which in low-ionization lines is dominated
by linear bar-like features \citep{ode00}, of which the Bright Bar (in
the lower-left of the images) is merely the most prominent example
(see \citetalias{gar06} for a more detailed discussion). In this paper
we discuss only the data around HH~528 out of the rich body of
information of this data-set because this is the lowest ionization
object among the many HH objects in the Orion Nebula.

\subsubsection{Radial Velocity Measurements of HH 528's Jet}

HH~528 is a long, collimated outflow along a position angle $\simeq
150\arcdeg$, which originates around 30\arcsec{} south of the
Trapezium as a broad jet and ends in a diffuse bowshock where it
interacts with the Bright Bar.

The \HH{528} jet can be best seen in our channel map centered on
$\Vhel = 10~\kms{}$ (Figure~\ref{fig:isomaps}) as an elongated
($20\arcsec \times 7\arcsec$) pink-orange region, centered on
\xy{0}{-72}.  It can also be seen at more blueshifted velocities, where
it is superimposed on the higher ionization Big Arc South, and more
redshifted velocities, where it is seen to be cut in two by a dark
extinction feature.

In \OI{}, the jet appears to have two components: a spatially broad,
smooth component, which is also seen in \SII{}, plus a series of
compact knots along its eastern side at \xy{-2}{-67}, \xy{-4}{-71},
and \xy{-3}{-75}. These knots are more blueshifted than the smooth
component, being most visible in the $+6~\kms{}$ channel. Further
south, we can identify two more knots of blueshifted emission in both
\SII{} and \OI{} at \xy{-18}{-92} and \xy{-14}{-89}, which show the
highest contrast against the surrounding nebula in the $+14~\kms{}$
channel. No \SIII{} emission is seen from the jet.

%%
%% Original version
%%
% The visibility of the jet in a given channel is strongly affected by
% the brightness and velocity of the surrounding nebular emission and so
% is not a reliable guide to the jet's radial velocity. Therefore, we
% have fitted Gaussian profiles to extracted one dimensional \OI{}
% spectra of the jet and adjoining background regions. We employed three
% different techniques: (1) fitting a double Gaussian to the jet+nebula
% sample; (2) same as (1) but constraining the velocity of the red
% component to be equal to the average velocity found by fitting a
% single Gaussian to the profiles from the two background nebula
% samples, and (3) fitting a single Gaussian to the result of
% subtracting the average background nebula profile from that of the
% jet+nebula. The derived velocities are $18.0$, $16.7$, and $16.0$,
% respectively, from the 3 methods, so we adopt an average value of
% $\Vhel(\OI) = +17\pm1~\kms$ for the jet.

%%
%% New version 23 Dec 2006
%%
\begin{figure}
  \centering
  \includegraphics{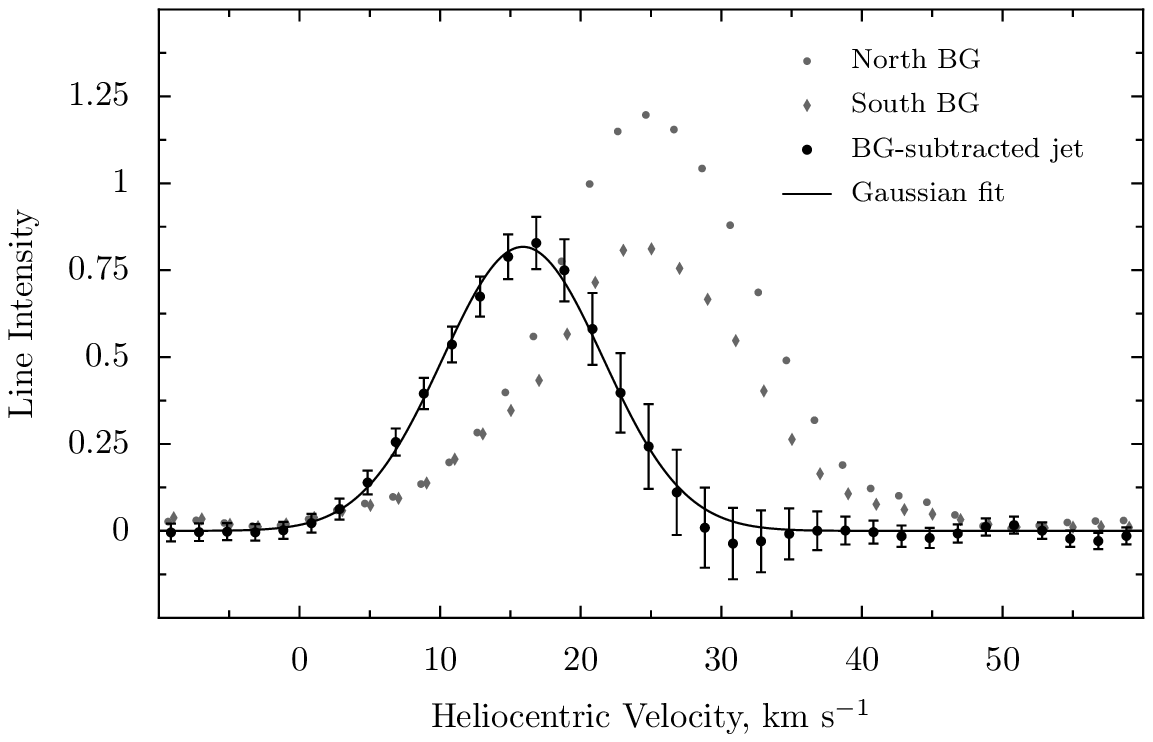} % hh528-jet
  \caption{Background-subtracted profile of the \HH{528} jet in the
    \OI{} line (black symbols with error bars), together with adjacent
    nebular background profiles (gray symbols). The continuous line
    shows a Gaussian fit to the jet profile. The error bars in the jet
    profile are a combination of the observational noise with the
    uncertainty in the estimated nebular background at the jet
    position (see text).}
  \label{fig:jetfit}
\end{figure}
The visibility of the jet in a given channel is strongly affected by
the brightness and velocity of the surrounding nebular emission and so
is not a reliable guide to the jet's radial velocity. Therefore, we
have extracted one dimensional \OI{} spectra from 5\arcsec{}-long
samples of the slit, which cover the jet and two adjoining background
regions. We have fitted a Gaussian profile to the
background-subtracted jet profile, assuming that the nebular emission
in the jet sample can be linearly interpolated from the two background
samples (see Fig.~\ref{fig:jetfit}). Although the peak velocity and
shape of the background nebular line is relatively stable in the
vicinity of the jet, its brightness shows substantial variation along
the slit, and this is the primary contribution to the uncertainty in
the jet profile on its redward side (we have used a conservative
estimate of one half of the difference between the two background
samples). The derived jet velocity from the Gaussian fit is
$\Vhel(\OI) = +15.9\pm1.8~\kms$, where the uncertainty is given by the
formal standard error in the fitting parameters.

We followed a similar procedure for \SII{} but this was complicated by
the presence of blueshifted emission from the Big Arc at the same
position as the jet. Nonetheless, we determine a jet velocity of
$\Vhel(\SII) = +16\pm3~\kms$, which is consistent with the more
precise \OI{} value.  The electron density in the jet is determined to
be $1900 \pm 500~\pcc$, which is very similar to the value in the
surrounding nebula.

\subsubsection{Radial Velocity Measurements of HH 528's Bowshock}
\begin{figure}
  \centering
  \includegraphics{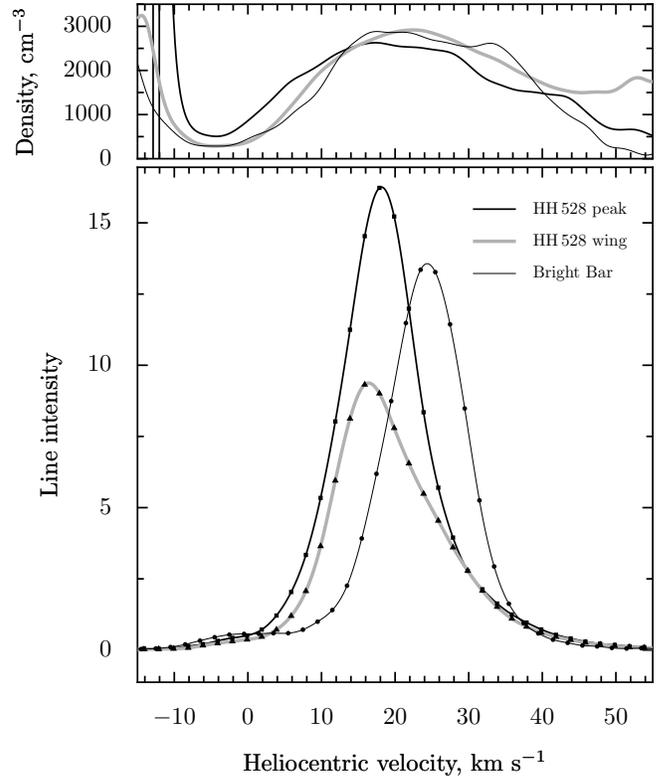} % hh528bow
  \caption[]{\SII{} line profiles of the \HH{528} bowshock from a
    70~$\mu$m slit (instrumental width: 6~\kms{}). The top panel shows
    electron densities calculated from the doublet ratio, while the
    bottom panel shows the intensity of the longer component.}
  \label{fig:hh528bow}
\end{figure}

The terminal bowshock of HH~528 is also visible as a blue-shifted
feature in our maps, showing the greatest contrast in the same
channels as for the jet. Although it is strongest in the
low-ionization lines, it is also present in \SIII{}. The brightest
part of the bowshock is a diffuse clump, of roughly 10\arcsec{} in
diameter, centered on \xy{-25}{-105}. It is displaced about
10\arcsec{} towards the Trapezium from where one would expect to find
the Bright Bar, assuming that the northeastern portion continued in a
straight line to the southwest. There is also blueshifted emission,
strongest in \OI{}, extending about 40\arcsec{} southwest along the
Bar, which seems to be a wing of the bowshock. Weaker blueshifted
emission also extends a similar distance to the east and north east,
which may represent the opposite wing of the bowshock.

Figure~\ref{fig:hh528bow} shows profiles of the \SIIlam{} line from
two positions in the bowshock and for an adjoining undisturbed region
of the Bright Bar at \xy{-65}{-90}. The bulk of the bowshock emission
is blueshifted with respect to the systemic velocity, peaking at
$\Vhel = +16$ to $+18~\kms$, as opposed to $+24~\kms{}$ for the
undisturbed bar. Note that it is impossible to reliably subtract the
background nebular emission from the bowshock profile, as was done for
the jet profile above. This is because the nebular emission shows
extreme brightness fluctuations that preclude the possibility of
interpolating its value from adjacent regions. In particular, the
emission from the Bright Bar is significantly depressed at the
position of the bowshock, as can be seen from
Figure~\ref{fig:hh528bow}. On the other hand, the bowshock component
is so strong that its velocity can be reliably estimated even without
background subtraction.

In addition to the blueshifted peak, the bowshock profiles also show
weak red wings, extending out to $+50~\kms{}$, which are lacking in
the Bar profile. The weak low density component around $0~\kms{}$ is
present in all the profiles and is an unrelated general feature of
this part of the nebula \citepalias{gar06}. The surface brightness of
the bowshock emission is very similar to that of the rest of the bar,
as is the electron density (upper panel of Figure~\ref{fig:hh528bow})
at 2500 to 3000~\pcc{}. The bowshock also shows the same ionization
stratification as the bar, with the emission from more highly ionized
lines becoming progressively more diffuse and shifted towards the
direction of the Trapezium. On the other hand, the \OI{} emission is
not as thin as in the Bar proper.

Roughly 15\arcsec{} to the south of the bowshock lies a curious zone
of low-ionization, low-brightness, scalloped emission features which
are most visible in the velocity range $+25$ to $+40$~\kms{}. Electron
density channel maps \citepalias{gar06} show a narrow ridge of
enhanced density that is continuous between these features and the
eastern portion of the Bright Bar.

\subsection{An HST Survey of the full Orion Nebula}

The new Hubble Space Telescope (HST) images used in this study were
obtained as part of program GO 10246, with Massimo Robberto serving as
the Principal Investigator.  This program imaged the Huygens region of
the Orion Nebula and its periphery using 104 pointings. The Advanced
Camera for Surveys (ACS) was the primary instrument and the pointings
were chosen so as to provide a continuous mosaic across the
nebula. Parallel observations were made with the Wide Field and
Planetary Camera (WFPC2), and the NICMOS infrared instrument, although
those data are not used in the current study.  The primary scientific
thrust of program GO 10246 is to study the properties of the Orion
Nebula Cluster stars and this drove the selection of the filters
employed; however, a complete mosaic was made with the F658N filter
(characteristic exposure time 340 s), which is equally sensitive to
both hydrogen H$\alpha$ (6563 \AA) and [N~II] (6583 \AA) emission. The
other filters employed and their characteristic exposure times were
F435W (420 s), F555W (385 s), F775W (385 s), and F850LP (385 s). The
observations were made in two observing campaigns, in October, 2004
and April, 2005. The wealth of images were combined and rendered into
a seamless mosaic.

\begin{figure*}
  \centering
  \includegraphics{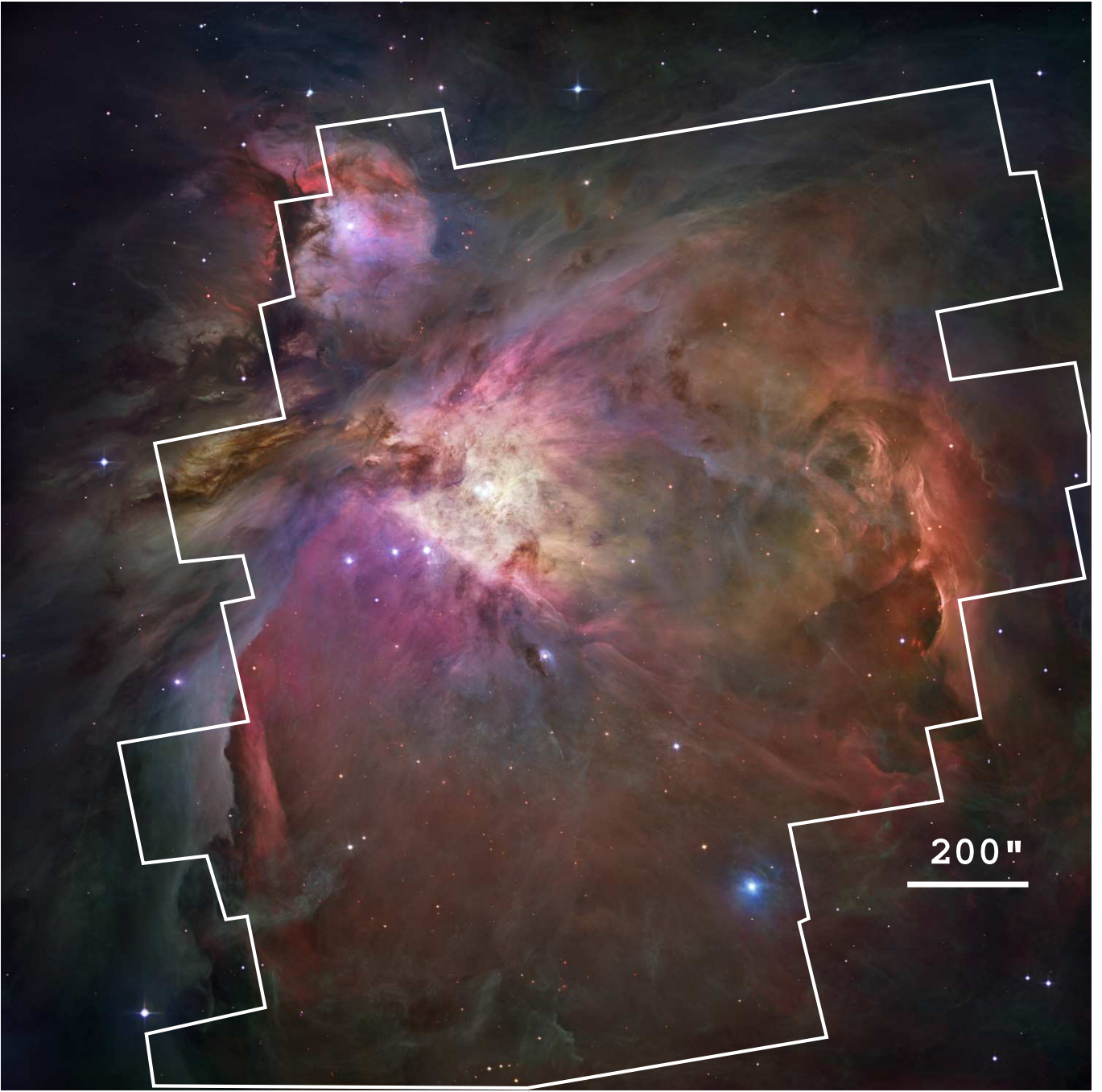} % Orion051209s10
  \caption[]{The core of this $30\arcmin \times 30\arcmin$ image of
    the Orion Nebula was prepared from HST ACS images. The irregular
    outline shows most of the field covered by the ACS survey
    indicated in Figure~\protect\ref{fig:bob3}.  Outside the
    boundaries of this survey, groundbased images were used.  The
    color coding is described in the text.}
  \label{fig:bob1}
\end{figure*}

The resulting picture from this survey is shown in
Figure~\ref{fig:bob1}, which is the combined image from all the
filters, including those dominated by continuum, rather than emission
lines. The outline of the field covered by the survey is shown in
Figure~\ref{fig:bob3} and the outer parts of this image have been
filled-in by images made by Massimo Robberto with the ESO La Silla 2.2
meter telescope. The color coding is red (F850LP+F775W, dominated by
scattered starlight with some [S~III] emission), red/orange (F658N,
dominated by \Halpha{}+[N~II]), green (F555W, dominated by [O~III] and
scattered starlight), and blue (F435W, dominated about equally by
hydrogen Balmer lines and scattered starlight).

\section{Properties of the HH Objects Associated with the Orion-S
  Region}

There are two regions of large scale outflows in the Orion Nebula: one
centered about 55\arcsec\ to the southwest from the brightest
Trapezium star (\thC) and the other centered on the BN-KL sources to
the northwest.  The two regions are clearly independent of one another
and in this paper we'll address only the outflows associated with the
southwest Orion-S region. As we shall see, it is likely that there are
multiple sources in this region. In this section we will summarize the
characteristics of the optical outflows that appear to be associated
with Orion-S and in the next section investigate whether there are
common sources for them.

For each object we will identify the discovery paper (when possible,
sometimes there were multiple steps in recognition of the nature of
the object), the values of the most accurate tangential and radial
velocities, derived velocity vectors, and any major papers that
discuss the object.  Since we are interested in the motions relative
to the host region, we convert heliocentric radial velocities to
velocities relative to Orion (\Vomc) by assuming the velocity of Orion
Molecular Cloud to be 28 \kms \citep{ode01}.  When motions in the
plane of the sky (tangential motions, \Vt) are available, these can be
combined with the \Vomc\ values to give spatial motions (\Vs) and in
all cases we will report the angle ($\theta$) of that velocity vector
with respect to the plane of the sky (positive values are towards the
observer) and the direction by the Position Angle (PA) measured
counter clockwise with respect to north. It is worth noting that
referring the angle of the velocity vector to the plane of the sky is
in contrast with earlier work (DOH04), where the orientation was given
with respect to the line of sight. The objects will be summarized in
order of their PA as viewed from Orion-S.

\begin{figure}
  \centering
  \includegraphics{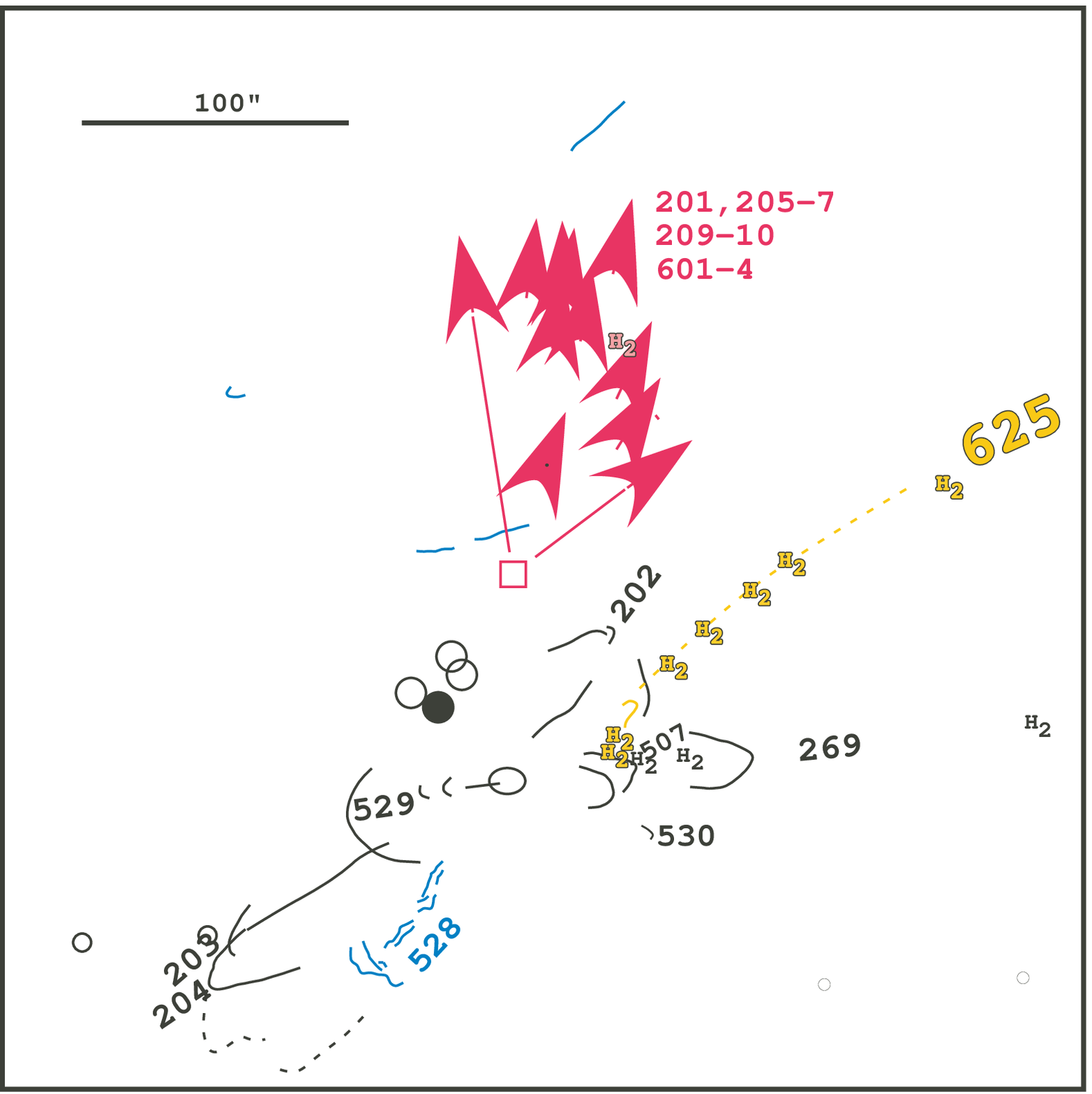} % CartoonCentralOutflows
  \caption[]{This drawing of a 450\arcsec\ square field in the center of
    the Orion Nebula depicts only the brighter stars and the HH
    objects. The coding of the symbols, letters, and lines is
    explained in the text. The dominant ionizing star \thC\ is shown
    as a filled circle.}
  \label{fig:bob2}
\end{figure}

\begin{figure}
  \centering
  \includegraphics{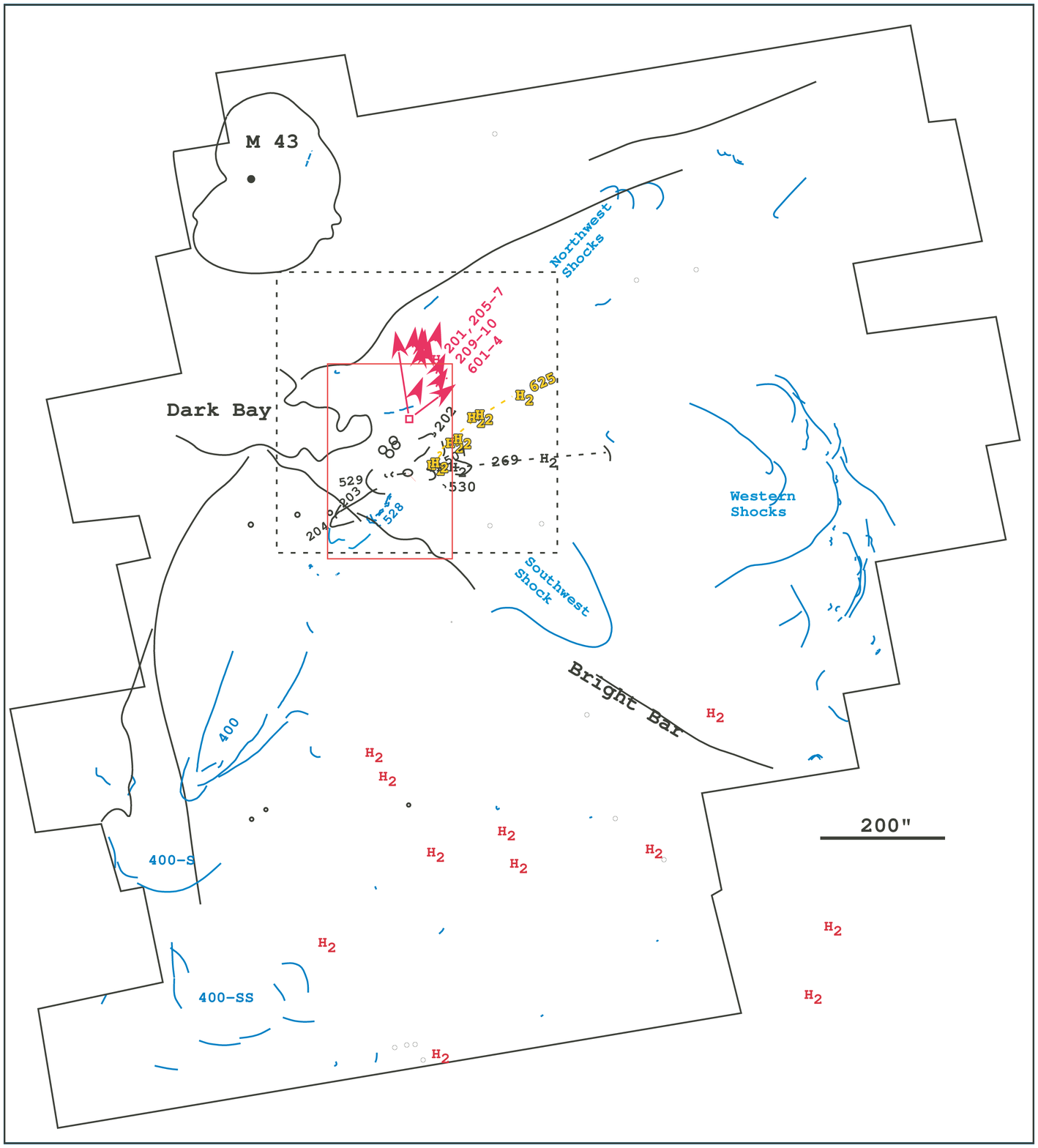} % CartoonLargeScaleFlows
  \caption[]{This drawing of a $1650\arcsec \times 1825\arcsec$ field
    around the Orion Nebula shows the HH objects in the outer parts of
    the nebula. The segmented line depicts the boundary of the survey
    made with the HST ACS, which has led to the discovery of many of
    these shocks. The code of the symbols, letters, and lines is
    explained in the text. The dashed box shows the central field
    shown in Fig.~\ref{fig:bob2}, while the red rectangular box shows
    the region mapped spectroscopically that is shown in
    Fig.~\ref{fig:isomaps}. }
  \label{fig:bob3}
\end{figure}

The forms and locations of the objects discussed are presented in
Figures~\ref{fig:bob2} and~\ref{fig:bob3}. The richness of features in
the Orion Nebula means that drawings like this are the best method for
isolating objects within a given class. Figure~\ref{fig:bob2} shows a
450\arcsec\ square field near the Trapezium stars (group of four
circles). Numbers indicate the HH object number. The red arrows and
open square indicate the HH objects associated with deeply embedded
BN-KL region. The \htwo\ symbols indicate the position of compact
\htwo\ objects in the survey of \citet{sta02}. The dashed yellow line
traces the path of the HH 625 components and the dashed black line
traces the faint outer features beyond HH 203-HH 204-HH 528. The open
ellipse is the position of the Optical Outflow Source (OOS) identified
by \citet[henceforth OD03]{ode03}.\defcitealias{ode03}{OD03}
Figure~\ref{fig:bob3} depicts a $1650\arcsec \times 1825\arcsec$ field
around the Orion Nebula. The irregular outer boundary shows the edges
of the coverage with the HST ACS survey and the dashed box the field
covered in Figure~\ref{fig:bob2}.  A few major features within the
nebula are labeled and the symbols employed are the same as in
Figure~\ref{fig:bob2}. Also shown are outer shock features that are
newly discovered in the ACS survey and which are discussed in more
detail in Section~5.4 below. In this case the features outside HH
203-HH 204-HH 528 are drawn as solid blue lines for clarity.

\subsection{HH 529}
The high ionization shock features now called HH 529 were visible in
the first HST images \citep{owo96}, their similar and high velocities
were noted in a low velocity resolution study by O'Dell et~al. (1997),
and they were designated as a single system by \citetalias{bom00}. The
shocks are easily visible in \OIII{} and \Halpha, but are faint in
\NII{}. They become broader and larger in the eastward direction of
apparent propagation.  The fact that they are of lower ionization
towards the west is discussed in detail by \citet{bla06}, who also
note that there are errors in the \OI{} data in \citetalias{bom00}'s
Figure~6. The emission in the filter meant to isolate the \OI{} line
at 6300~\AA\ is most likely dominated by the \SIII{} 6312~\AA\ line
and the spectrum given for \OI{} is an accidental reproduction of the
\OIII{} spectrum.  The interpretation that these are shocks located
within the ionized layer of the Orion Nebula \citep{ode97a} is
strengthened from the study of their spectra \citep{bla06}.
Components of this system were originally identified by their high
radial velocities \citep{lee69,cas88,mas95} and \citetalias{bom00}
measured some of the components with the Keck HIRES with good spatial
and velocity resolution.  \citetalias{doi04} mapped the entire object
in \Halpha, \OIII{}, and \NII{} in their slit study of the Huygens
region of the Orion Nebula.  Because of the multiple components of
this HH object, we will designate each according to the position-based
system for the Orion Nebula introduced by \citet{ode94}. We will adopt
the radial velocities from \citetalias{doi04} and the tangential
velocities of \citetalias{ode03} as they have a longer time-base than
\citet[henceforth DOH02]{doi02}\defcitealias{doi02}{DOH02} and
\citetalias{bom00}. The heliocentric velocities of the features near
\OW{150}{352}, \OW{158}{353}, and \OW{165}{358} are $-44\pm 4 \ \kms$,
$-35\pm 9\ \kms$, and $-26\pm2\ \kms$ respectively, which yields
values of $\Vomc= -72$, $-63$, and $-54\ \kms$. The \Vt\ values of
these features are $88\pm20\ \kms$, $46\pm16\ \kms$, and $101\pm16\
\kms$ towards PA values of $98\arcdeg\pm8\arcdeg$, $110\arcdeg\pm23
\arcdeg$, and $96\arcdeg\pm26\arcdeg$. The resultant values of \Vs\
are $114\ \kms$, $78 \ \kms$, and $115\ \kms$, with $\theta=
39\arcdeg$, $54\arcdeg$, and $29\arcdeg$. Since the errors in the
velocities that are used to derive $\theta$ are significant and the
calculated angles are correspondingly uncertain, it is not clear if
these three features are moving at different angles or along a common
average angle of about 41\arcdeg. In the discussion in \S~4, we shall
assume the latter.

\subsection{HH 203}

HH 203 was first recorded by M\"unch \& Wilson (1962) and has been the
subject of numerous subsequent studies of its appearance in different
ions \citep{ode97b}, its electron density (Walsh 1982), radial
velocity (O'Dell, Wen, \&\ Hu 1993, DOH04), and tangential motions
(Cudworth \&\ Stone 1977, Hu 1996, DOH02).  Because of its proximity
to \thA\ M\"unch \&\ Taylor (1974) interpreted the object as the
result of the interaction of a stellar wind with the ambient nebular
gas, but the symmetry of the shock and the presence of a high velocity
jet leading to it (O'Dell et~al. 1997a, Rosado et al. 2001, DOH04)
indicate that it is a jet driven shock. Henney (1996) argues that its
asymetry in brightness can be explained by the jet striking denser
ambient material at grazing incidence.  The tip of this feature is a
series of well defined shocks. Originally, O'Dell et~al. (1997b)
argued that the shock is formed when the driving jet strikes the
foreground Veil of neutral material that lies in front of the nebula
\citep{ode01}, but a more accurate knowledge of the position of the
Veil \citep{abe04} and an improved knowledge of the shock's trajectory
allowed DOH04 to establish that HH~203 arises where the driving jet
strikes denser nebular material where the ionization front of the
Orion Nebula tilts up. This tilt is what gives rise to the Bright Bar
feature that runs from northeast to southwest, passing near \thA.

The radial velocity map of DOH04 yields a heliocentric velocity for
HH~203 of $-46 \pm 15 \ \kms$, which means $\Vomc= -74\ \kms$.  The
tangential velocity study of DOH02 gives a value of $\Vt = 73 \pm23\ 
\kms$ towards $\mathrm{PA}=134\pm17\arcdeg$.  This means that the spatial
velocity is $\Vs =104\ \kms$ and $\theta=45\arcdeg$.

\subsection{HH 204}

HH~204 is close in position to HH~203 and has been studied at the same
time as HH~203 in the previous investigations. It extends slightly
further from the center of the Orion Nebula and is quite different in
appearance, the head of the parabolic shock form being highly
flocculent in appearance, rather than having the multiple small shocks
present in HH~203. It has extended [O~III] emission within the
parabolic shock form, which is a strong indication that it is an
object that is being photoionized, possibly either by \thA\ or more
likely by \thC. The presence of low ionization lines in its tip
indicate that it too is striking a dense, low ionization region of the
nebula, like HH~203, or, alternatively, that the compression of the
ionized gas by the shock leads to the trapping of the ionization
front.

The radial velocity map of DOH04 yields a heliocentric velocity for HH
204 of $-18 \pm 18 \ \kms$, which means $\Vomc= -46\ \kms$.  The
tangential velocity study of DOH02 gives a value of $\Vt = 92 \pm 10 \
\kms$ towards $\mathrm{PA}=137\arcdeg\pm7\arcdeg$.  This means that
the spatial velocity is $\Vs =103\ \kms$ and $\theta=27\arcdeg$.

\subsection{HH 528}

Tangential velocities in the HH~528 jet have been measured for
individual knots in the \OI{} (BOM) and \SII{} (DOH02) lines.  Three
of these coincide with knots for which we have measured radial
velocities. Taking the vector mean of the proper motion measurements
of the knots gives a tangential velocity of $\Vt = 25 \pm 6~\kms$ at a
position angle of $\mathrm{PA} = 139\pm15\arcdeg$, which is consistent
with the orientation of the line between the jet and the bowshock
($\mathrm{PA} = 148 \pm 5\arcdeg$) and the major axis of the jet as
measured from our \OI{} channel maps ($\mathrm{PA} = 150 \pm
10\arcdeg$). Our fits to the jet component in the \OI{} and \SII{}
line profiles give $\Vhel = 17 \pm 2~\kms{}$, implying that $\Vomc= 11
\pm 3~\kms{}$, in which the uncertainty includes the effect of the
2~\kms{} velocity dispersion of stars in the Trapezium cluster (Jones
\& Walker 1988). Therefore $\theta =24\arcdeg \pm 8\arcdeg$ for the
jet.

Tangential velocities have also been measured for knots in the
bowshock of HH~528 \citepalias{bom00,doi02}, although the dispersion
in magnitude and direction is rather large. Taking the vector mean of
the 6 measured knots gives $\Vt= 21 \pm 8~\kms{}$ at a PA of $206\pm
24~\arcdeg$. The highest velocity ($33~\kms$ at $\mathrm{PA} =
176\arcdeg$) is seen in knot \OW{182}{513} at \xy{-25}{-109}, which
lies at the very nose of the bowshock. Our \SII{} spectra of the
bowshock (Figure~\ref{fig:hh528bow}) show peak velocities of $\Vhel =
16$--18~\kms{}, very similar to that of the jet, implying $\Vomc \sim
11~\kms{}$.  Given the uncertainties involved, the data are therefore
consistent with the jet and bowshock sharing the same space
velocity.\footnote{Note that our results are inconsistent with the
  claim of \citet{smi04}, based on unpublished Fabry-Perot
  observations, that HH~528 is redshifted. }
%  of $\simeq
% 27~\kms{}$ in a direction of $\mathrm{PA} \simeq 145\arcdeg$, inclined
% at a small angle ($\simeq 20\arcdeg$) from the plane of the sky
% towards the observer.

\subsection{Features possibly related to HH 203-204-528}

% XX Will. I can't see it very well on the 10246 F658N filter image,
% only slightly better on the color mosaic, and only marginally on the
% old masterfXXX.hhh images. If we can't come up with a good
% illustration, I would say that we leave it out except to mention it in
% the discussion session.

% XX Will.Make a figure that shows these(?).

% XX Bob: Is this the figure (Fig.~\ref{fig:large-bow}) that you mean?
% It might be best cropped horizontally and maybe shown in two versions,
% one with an explanatory overlay. I will also remove the little green
% boxes.

\begin{figure*}
  \centering
  \includegraphics[width=0.95\linewidth]{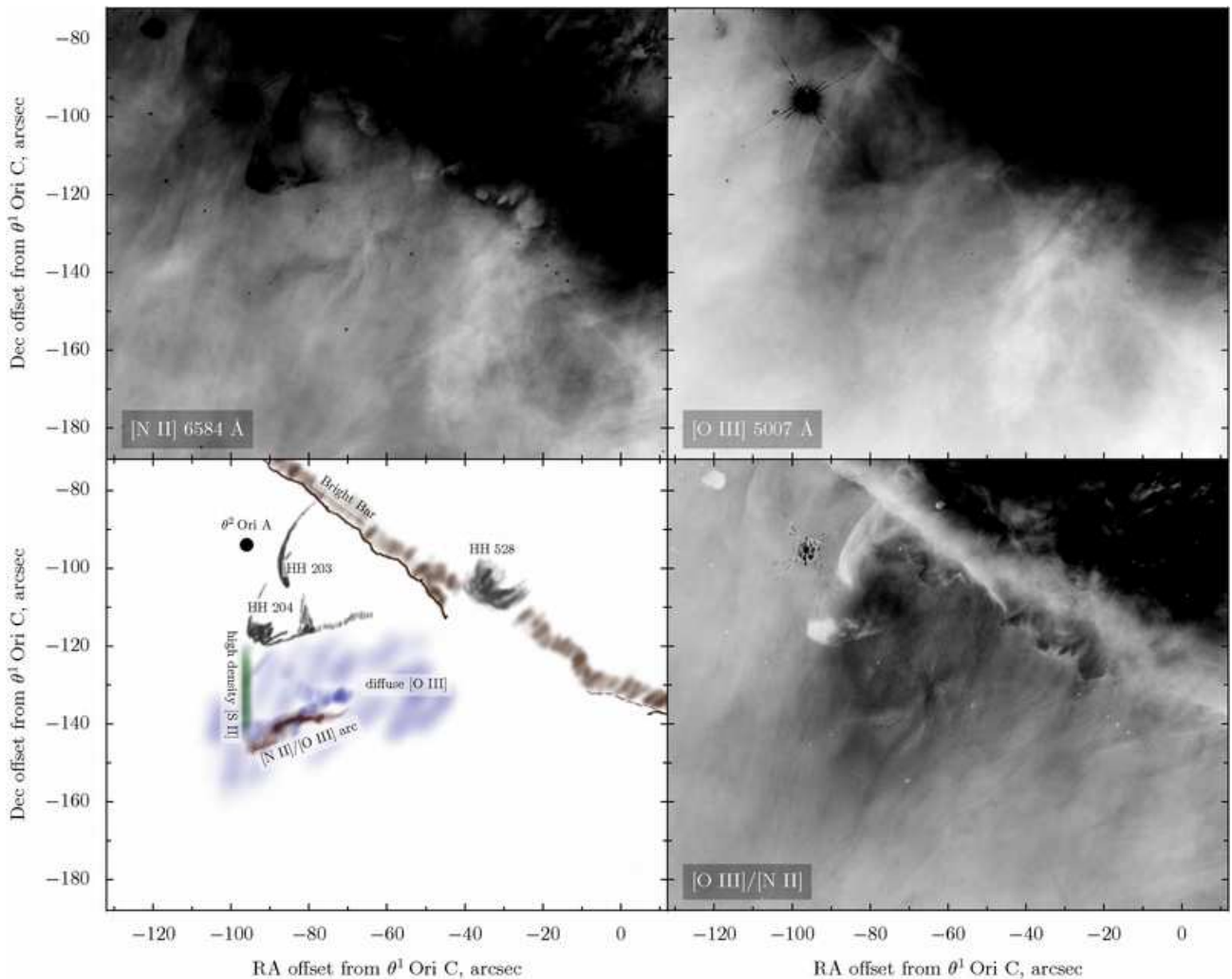} % BigBow-graph
  \caption[]{Deep \textit{HST} WFPC2 images of the region south-east
    of the Bright Bar in \NII{} (upper-left) and \OIII{}
    (upper-right). These are shown as negative grayscales that have
    been tuned to emphasize faint features, with the result that the
    emission from the inner nebula is saturated. Also shown is the
    \OIII{}/\NII{} ratio (lower-right, with black signifying higher
    ionization), and a cartoon (lower-left) showing the principal
    features.}
  \label{fig:large-bow}
\end{figure*}

Figure~\ref{fig:large-bow} shows faint, high-ionization emission to
the south-east of the Bright Bar. The morphology of the emission is
reminiscent of a bowshock shape, which may represent an outer shock
related to the HH~203/204 system. Alternatively, we cannot rule out
that it may be associated with HH~528. Superimposed on this feature is
a faint arc of relatively high-density gas, identified in \SII{}
channel maps by GH07.

% XXX This section still needs to be expanded slightly XXX

\subsection{HH 400}

HH 400 was discovered by \citet{bal01} on \Halpha{} and \SII{} images
of the Orion Nebula and they also report on velocity mapping of the
southern part of the nebula in the \SII{} lines. This is the largest
of the known HH objects in the Orion Nebula, being a group of large
shocks with the furthest tip extending 633\arcsec\ from Orion-S with
$\mathrm{PA} = 146\arcdeg$. In Figures~\ref{fig:bob2} and
\ref{fig:bob3} we see that there are even larger shocks extending to
789\arcsec\ at $\mathrm{PA}=147\arcdeg$ and the shock features at
970\arcsec\ at $\mathrm{PA}=160\arcdeg$ may also be associated if the
curvature continues. \citet{bal01} give only radial velocities
relative to the local region of the nebula, finding it redshifted by
$8$--$20\ \kms$. Since there are no published \SII{} velocities for
this part of the nebula, and there are no tangential velocities for
HH~400, it is impossible to derive its spatial motion. However,
\citet{bal01} argue from the shape of the HH~400 shocks that it is
moving almost in the plane of the sky.

\subsection{HH 530}
HH 530 was first identified in \citetalias{bom00}, where it is depicted
in their Figure 22. \citetalias{ode03} improved the measurements of the
tangential motions and questioned whether the shock at \OW{108}{430}
is part of this outflow (since it may be associated with the silhouette
proplyd \OW{114}{426}).  OD03 measured the components \OW{105}{416}
and \OW{105}{417}, which are parts of a well defined shock oriented
almost due west. They have an average motion of $\Vt = 28 \pm 5 \
\kms$ towards $\mathrm{PA}=244\arcdeg \pm26\arcdeg$. There is a nearby
3\arcsec\ feature oriented almost east-west ($\mathrm{PA}=260\arcdeg$)
and may be an associated jet. This latter feature was found to be
moving at $\Vt =63 \pm 7 \ \kms$ towards $\mathrm{PA}=255\arcdeg
\pm 4\arcdeg$. There are no published radial velocities of this
outflow.

\subsection{HH 269}
HH 269 was first identified as an HH object in \citep{wal95},
following earlier observations of its peculiar form \citep{fib76} and
density enhancements in this region \citep{wal94}.  It is composed of
a series of shock structures oriented towards the west.  Although some
[O~III] emission is seen, the object is brightest in low ionization
lines \citep{wal95} and there are associated \htwo\ features in the
images of \citet{sta02}.  The tangential velocities of the tip of the
east shock were measured by OD03 to be $40 \pm 26 \ \kms$ towards
$\mathrm{PA}=283\arcdeg\pm12\arcdeg$. The heliocentric radial velocity
of the east shock is \citep{wal95} $13\ \kms$ ($\Vomc=-15\
\kms$). These values yield $\Vs=43\ \kms$ with $\theta=21\arcdeg$.

\subsection{HH 507}
HH 507 was first identified in \citetalias{bom00}, where it is depicted
well in their Figures 22 and 23. It lies along the axis of HH 269, but
is clearly separate as it has its own characteristic velocities while
features to the east and west have the common motion of HH~269.  It
has the form of a west-northwest oriented parabola. \citetalias{ode03}
found that $\Vt = 27\pm12\ \kms$ towards $\mathrm{PA}=298\arcdeg
\pm41\arcdeg$. \citetalias{bom00} point out that there is a linear
feature along the symmetry axis of this shock and has
$\mathrm{PA}=296\arcdeg$. If associated with HH 507, this means that
the driving jet is not the nearby proplyd \OW{117}{352} as it is not
aligned with that object. BOM hypothesize that HH~507 is driven by
outflow from source FIR~4 \citep{mez90}, although a more complete
discussion is presented in our \S~5. There are no published radial
velocities of HH~507.

\subsection{HH 625}

HH 625 was identified in \citetalias{ode03} and is unlike any of the
other HH objects in the Orion Nebula. It has a peculiar dark feature
lying 76\arcsec\ at $\mathrm{PA}=267\arcdeg$ from \thC\ and that
feature has a bright rim on its northwest boundary, the feature being
oriented towards $\mathrm{PA}=325\arcdeg$. \citetalias{ode03} measured
the tangential velocity of the associated bright features to be moving
at $\Vt =26 \pm 4 \ \kms$ towards $\mathrm{PA}=304\arcdeg
\pm12\arcdeg$. There are associated \htwo\ features seen in Subaru
images \citep{2000PASJ...52....1K} and in the \htwo\ survey of the
region by \citet{sta02}. The small \htwo\ knots seen in the latter
study begin with a linear alignment close to $\mathrm{PA}=313\arcdeg$,
then curve slightly clockwise, reaching a distance of about
204\arcsec\ from their point of origin that is discussed in \S~5.

There are no published radial velocities of the optical features in
this outflow. However, if one adopts a value of $\Vomc=80\ \kms$, as
suggested by the CO molecular outflow \citep{zap05} that is the
driving source of this HH object (as discussed in \S~5), then $\Vs
=84\ \kms$ and this part of the object has $\theta=73\arcdeg$.

\subsection{HH 202}

HH 202 was one of the first HH objects identified in the Orion Nebula
\citep{can80}. The best imaging study was carried out with the HST
\citep{ode97b}, where it was revealed that there are a series of fine
scale structures near the tip of a broad parabolic
envelope. Tangential velocities have been measured by \citet{cud77},
\citetalias{doi02}, and \citetalias{ode03}. Radial velocities have been
measured by \citet{mea86}, \citet{cla94}, \citet{ode91}, and
\citetalias{doi04}. The northwest tip of the object is extremely complex
in both images made in different ionic emission and in long-slit
spectroscopy. The brightest part is near this tip and is called
HH~202-S. The extended emission in \OIII{} within the parabolic
envelope argues that like HH~204, this shock is photoionized from the
rear, in this case almost certainly by \thC.

DOH04 give a heliocentric velocity of $-39 \pm 2 \ \kms$ for the
bright HH~202-S complex, which corresponds to $\Vomc =-67\ \kms$, and
\citetalias{ode03} give $\Vt=53 \pm 14 \ \kms$ towards
$\mathrm{PA}=332\arcdeg\pm13\arcdeg$ for the same region. These values yield
$\Vs =85\ \kms$ and $\theta=52\arcdeg$.

\section{The Combined Properties of the HH Objects}

\begin{figure}
  \centering
  \includegraphics{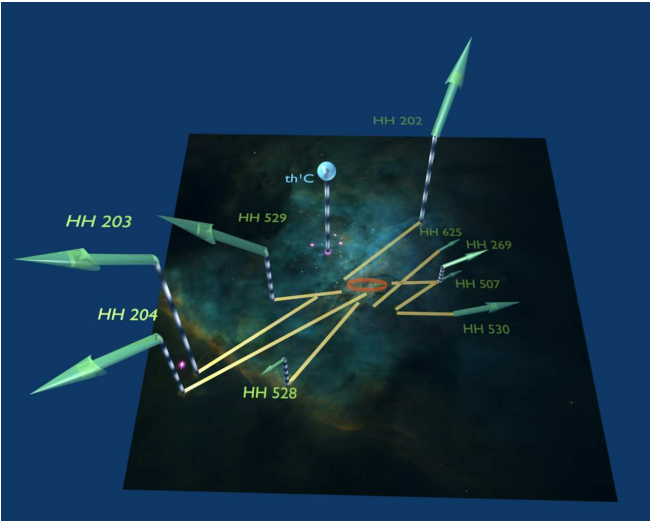} % bob-arrows6
  \caption[]{The motions of the cataloged HH objects in the Huygens
    region of the Orion Nebula are shown superimposed on an early
    mosaic of WFPC2 images \citep{owo96}.  The position of the
    dominant ionizing star (\thC) with respect to the ionization front
    is shown as a sphere while the location of O'Dell \&\ Doi's (2003)
    Optical Outflow Source is shown as an ellipse. The ionization
    front is actually an irregular, concave (as viewed by the
    observer) surface. The observer is located at an extension of the
    vertical axis in this figure. When both radial and tangential
    velocities are known, their velocity vectors are shown (all are
    blue-shifted) while those objects with only tangential velocities
    are drawn as being in the plane of the figure.  The lines drawn in
    the plane of the figure indicate the symmetry axis of each shock.}
  \label{fig:3d}
\end{figure}

In the case where we can determine spatial velocity vectors (PA, \Vt,
and $\theta$ are known) and have at least an approximate idea of the
place of origin, we can depict these flows in a manner that illuminates
what is happening. Figure~\ref{fig:3d} shows the central region of
the nebula in three dimensions. For simplicity, we have assumed in
calculating the position along the vertical axis that the objects all
originate from the center of the OOS. This assumption is certainly not
valid for all the HH objects, but affects only little the general
picture of what is going on. In each case the direction of the
velocity vector is that derived from \Vt\ and $\theta$. It should be
recalled that the inner region of the Orion Nebula has been mapped in
three dimensions by \citet{wen95}, who determined that the
surface is concave (as viewed by the observer), with \thC\ lying about
0.2~pc (90\arcsec) in front of the ionization front and there is a
local rise of about 0.05~pc (22\arcsec) at Orion-S. The last feature
indicates that Orion-S is a denser region within the host Orion
Molecular Cloud and that the ionization front has not penetrated as
far there.

The first striking thing about Figure~\ref{fig:3d} is that all of the
optical objects are blueshifted. This indicates that the sources for
these HH objects lie in the background of the visible nebula or even
in the optically obscured region beyond the IF. One would expect the
optical extinction to increase significantly as one begins to
penetrate the Photon Dominated Region (PDR) because the density jumps
there.  If a source were close to the IF, its redshifted component
would quickly disappear, unless the flow was diverted into being
blueshifted.

The fact that several flows seem to be paired (e.g., HH 529 and HH
269, HH 203/204 and HH 202) has led multiple authors to posit that
these are bipolar flows from a single source, with the originally
redshifted component being altered into having a blueshift without a
significant change in the orientation on the plane of the sky
\citep[OD03]{ode97a,bal01,ros01}. This argument is especially
attractive for the HH 529-HH 269 pairing because one sees the
components moving in opposite directions in the plane of the sky,
leaving a gap only in the OOS region \citepalias{ode03}.

A process for deflection of a redshifted component has been proposed
and modeled by \citet{can96}.  They establish theoretically that a jet
passing through a density gradient will be deflected towards the
direction of the lower density. This is extremely attractive for
explaining what we see in the Orion Nebula because we know that there
is a strong density gradient away from the IF, being lower towards the
observer \citep{ode01} and theoretical models indicate that the
density in the neutral PDR should rise as one passes the IF\@. Whether
or not a beam is deflected by or penetrates the PDR will be dependent
upon many factors, including geometry, magnitude of the density
gradient, mass flux of the jet, so that one cannot make an accurate
prediction of what will occur.

This means that redshifted flow from a source located within the
ionized zone of the nebula will eventually reach the IF. If it is
deflected sufficiently it can become a blueshifted flow and if it
passes through or remains in the PDR then it will be optically
invisible.

If the source is obscured by extinction (embedded), then the
conversion of a redshifted into a blueshifted component means that the
source must lie close enough behind the PDR so that there is still an
increase of density away from the observer. There is some evidence
that this is the case for the flow from the OOS source since
\citetalias{ode03} calculated that the source was only a few
hundredths of a parsec beyond the IF.

Unfortunately, the \citeauthor{can96} mechanism can also make it more
difficult to identify the location of a source.  This is because if a
jet strikes a region with a tilted density gradient not only will
there be the potential of changing a redshift into a blueshift, but,
the direction in the plane of the sky can be changed. However, one
expects that this alteration will be less important since the
concavity of the IF is not great except in the region of the Bright
Bar. This mechanism may be what produces the curvature of the path of
the components of HH~625, which \citetalias{ode03} argue is a flow
grazing along just behind the IF.

There is another mechanism that complicates identification of a source
of an outflow by extrapolating back in the plane of the sky and that
is deviations in the direction of the flow caused by a wind blowing on
the jet.  This mechanism has been invoked by \citet{mas01} to explain
the bent small-scale bipolar jets that one sees in the Orion Nebula,
the object associated with LL Orionis being the best example
(BOM). The side wind is most likely to be the flow of gas away from
the bright central region of the nebula. It may be that the well
defined curvature necessary to connect HH 203+HH 204 to HH 400 and its
southernmost components is caused by such a wind.

The second striking thing about Figure~\ref{fig:3d} is that the backwards
projection of the velocity or symmetry vectors cross in a relatively
small area, close to Orion-S. In \S~5 we will discuss the probable
sources that lie in this region, putting special emphasis on
identification of infrared or radio compact sources and the molecular
flows known to exist there.

\section{Discussion}

\subsection{Comparison of the Optical Objects with the Embedded
  Molecular Outflows}

\begin{figure}
  \centering
  \includegraphics{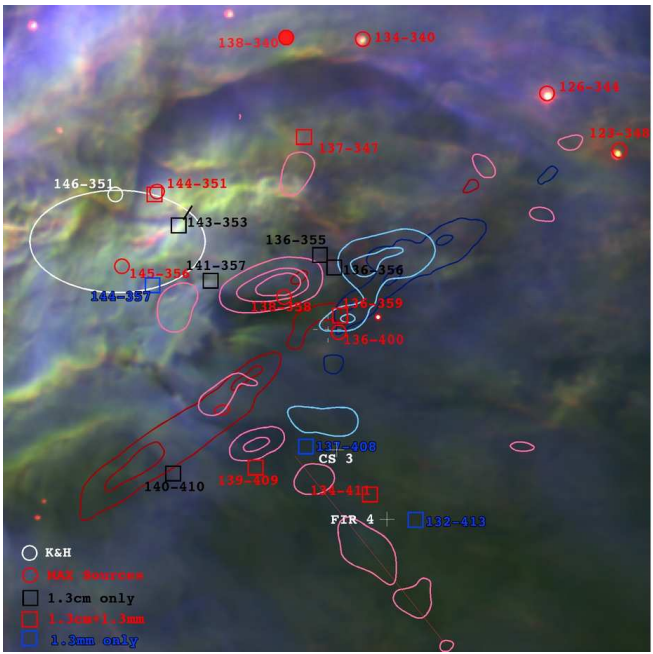} % smfov
  \caption[]{The CO (bright blue and red) and SiO (pink and pale-blue)
    emission in a 44\arcsec\ square region around Orion-S is shown
    superimposed on a composite of HST WFPC2 images as described in
    the text.  The white ellipse indicates the Optical Outflow Source
    region identified by OD03 as the region of the source of HH 269
    and HH 529.  The squares and circles represent the positions of
    various radio and infrared compact sources, with the white circle
    representing the 1.6 $\mu$m and 2.2 $\mu$m source from Hillenbrand
    \&\ Carpenter 2000, red circles the 10 and 20 $\mu$m MAX study of
    \citet{rob05} with the brightest source filled-in, the black
    squares the 1.3 cm sources of \citet{zap04}, the red squares the
    sources seen at both 1.3 cm and 1.3 mm, and the blue squares the
    sources seen only at 1.3 mm \citep{zap05}. FIR 4 is the 1.3 mm
    source of \citet{mez90} and CS~3 is from \citet{mun86}.}
  \label{fig:bob4}
\end{figure}
\begin{figure}
  \centering
  \includegraphics{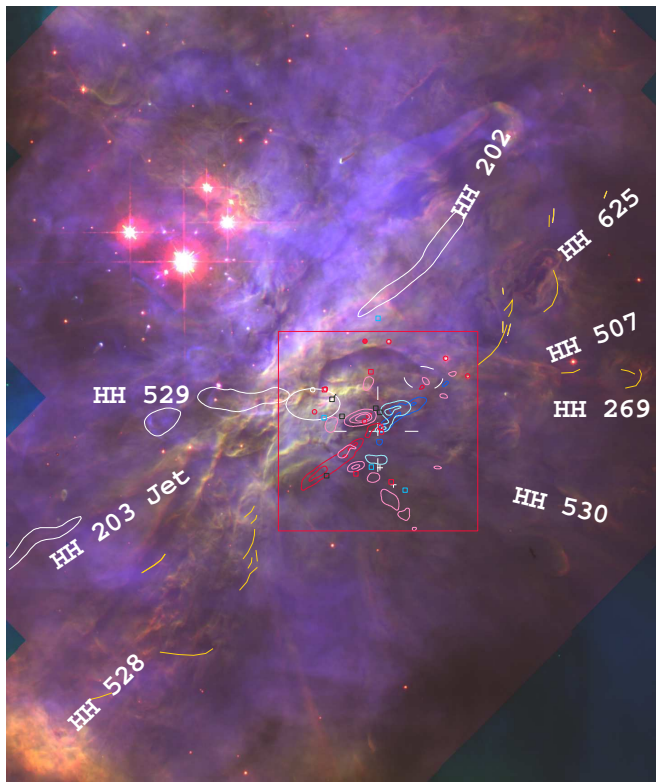} % lgfov
  \caption[]{The features of Figure~\ref{fig:bob4} (inner box) are shown
    on a $143\arcsec\times 168\arcsec$ field within the Orion Nebula
    compiled from HST WFPC2 with the same color coding as
    Figure~\ref{fig:bob4}. The field covered in Figure~\ref{fig:bob4}
    is outlined in red.  The location of various HH objects is
    indicated. Thin yellow lines indicate \htwo\ emission from the
    Subaru images (Kaifu et~al. 2002) and the contours indicate high
    velocity features delineated by enhanced HeI 10830 emission as
    described in the text.}
  \label{fig:bob5}
\end{figure}
Significant progress has been made recently in the understanding of
molecular outflow from the Orion-S region.  The most important recent
contributions are the studies of CO (Zapata et~al. 2005) and SiO
(Zapata et al. 2006), which exceed in sensitivity and angular
resolution the previous investigations of this region. These two
studies are summarized in Figures~\ref{fig:bob4} and~\ref{fig:bob5}.

In Figure~\ref{fig:bob4} the CO results are shown as contours of the
redshifted (red) and blueshifted (blue) components. For simplicity we
have drawn only the outermost brightness contours and a few of the
inner. \citet{zap05} show that the flow aligned along
$\mathrm{PA}=306\arcdeg$ reaches maximum values of $\pm80\ \kms$.  SiO
emission in the outflows from this region far exceed the normal
nebular abundance and must arise from destruction of grains within the
molecular cloud or the PDR lying immediately beyond the nebula's IF
(Zapata et~al. 2006).  The results of the SiO study are shown in
Figure~\ref{fig:bob4} in a similar manner to the CO results, except
that redshifted emission is depicted as pink and blueshifted as
pale-blue. The primary emission occurs along an axis of PA=284\arcdeg
and reaches velocities of \Vomc=-90 \kms\ and \Vomc=75 \kms.  These
radio results are shown displayed on an HST WFPC2 optical image with
the color coding [S~II]-red, [N~II]-green, and [O~III]-blue.  In
addition, we show the positions of compact radio and infrared sources
in the vicinity of Orion-S, as discussed later in this section, in
addition to the Optical Outflow Source of \citetalias{ode03}.

A wider field around Orion-S is shown in Figure~\ref{fig:bob5}, where
we have added to the inner region data of Figure~\ref{fig:bob4} the
outer optical objects (designated by their HH catalog numbers), and
\htwo\ emission indicated on the Subaru images
\citep{2000PASJ...52....1K}.  Highly blueshifted optical emission is
shown by closed contours of thin yellow lines. The location of the
compact and narrow high velocity components are best delineated in
\citetalias{doi04}, but we have shown the contours of \ion{He}{1}
emission from \citet{tak02}. This is useful because the \ion{He}{1}
10830~\AA{} line is optically thick because its lower state is
metastable. Differences of radial velocity shift the line outside the
core of the absorption and allow that radiation to selectively escape.

\subsection{A Well Defined Connection between the CO Outflow, its
  Infrared Source, and HH 625}

The new high spatial resolution CO images of the Orion-S region have
provided the necessary ingredients to firmly establish that HH 625 is
driven by the blueshifted outflow from the embedded source
\OW{136}{359} as previously suggested in \citetalias{ode03}. It is
likely that the beam of molecular gas remains very confined and is
grazing along the PDR, occasionally breaking through the irregular and
concave IF\@. In this way one can reconcile the CO, \htwo, and optical
observations.

It is natural to look for optical features corresponding to the
redshifted CO outflow.  The deflection of the beam that would be
necessary to produce an optical counterpart would also produce the
possibility of deviation in PA, thus complicating making an
identification.  The most promising candidate would be HH 528 because
its PA is closest to that of the redshifted CO component. There is
probably not a connection with HH 203 and HH 204 as their driving jet
is aligned towards the OOS and linking them with the CO outflow would
mean that the jet would have to suffer two deviations in PA, the first
to place it into alignment with the HH 203 jet, then a second to point
it towards the shocks we call HH 203 and HH 204.  The key element of
confirming an association of the redshifted CO outflow with HH 528
would be to discover tangential motions that connect the southeast end
of the molecular outflow with the low ionization features that form
the base of HH 528.

\subsection{A likely Connection between the SiO Molecular Outflow, its
  Infrared Source, and HH 507}

The well defined SiO outflow has a strong candidate for its source
(\OW{135}{356}, \citealp{zap06}), however, there no obvious
connections with optical HH objects. The blueshifted component that is
elongated towards $\mathrm{PA}=284\arcdeg$ does point towards HH~507,
which lies at $\mathrm{PA}=283\arcdeg$ from \OW{135}{356}. As noted in
\S~3.9, the shock forming HH 507 is oriented towards
$\mathrm{PA}=296\arcdeg$ and is moving towards $\mathrm{PA}=298\arcdeg
\pm41\arcdeg$. Moreover, as noted by \citetalias{bom00}, there is a
trailing linear feature, which may be a jet, that is oriented towards
$\mathrm{PA}=296\arcdeg$. These combined characteristics argue that HH
507 is associated with the blueshifted SiO outflow from
\OW{135}{356}. The discovery of the SiO outflow and its alignment with
HH~507 means that the conclusions presented here supercede those of
\citetalias{bom00}, where it was argued that HH~507 originated from
the region of CS~3 or FIR~4. The possibility of HH 269 being driven by
the blueshifted component of the SiO outflow is discussed in \S~5.5.
%% Changed to above WJH following CRO - 18 Aug 2006
% It is unlikely that the more extensive HH
% 269 features arise from either the SiO or CO outflows because the
% tangential velocity study of \citetalias{ode03} indicate that there
% are features with the same tangential velocities that extend east to
% the edge of the OOS ellipse in Figures~\ref{fig:bob4}
% and~\ref{fig:bob5}.

There are no obvious connections of the redshifted SiO outflow with
optical HH objects. Associations with any of HH~203, HH~204, or HH~528
could be made by a single deflection that resulted in both a change of
PA and radial velocity.  There is an incomplete shock lying 34\arcsec\
at $\mathrm{PA}=106\arcdeg$ from \OW{135}{356} which is oriented
towards $\mathrm{PA}=121\arcdeg$. The latter value is very close to
$\mathrm{PA}=124\arcdeg$ that applies to the HH~203 jet, which argues
that there is an association, if the deviation occurs at just about
the position where this shock appears.

\subsection{Possible Connections between Molecular Outflows, Infrared
  Sources, and Distant Shocks}

\citet{zap06} point out that sources CS~3 \citep{mun86} and FIR~4
\citep{mez90} are likely to be the same, a suggestion made previously
by \citep{gau98}, and that this source could be related to their
strong millimeter source \OW{137}{408}.  The outflow in the region of
this source is complex, there being a redshift-blueshift pair of SiO
features lying immediately to the north of it, and a series of
redshifted SiO features extending towards $\mathrm{PA} =
219\arcdeg$. \citet{sch90} find a redshifted CO outflow 120\arcsec\
long extending towards $\mathrm{PA} = 211\arcdeg$ originating from the
CS~3-FIR~4 region. \citet{zap06} argue that this implies that both the
SiO and CO outflow originate from close to \OW{137}{408}.  There are
no cataloged HH objects in the direction of the outflows away from
\OW{137}{408}. However, at a distance of 397\arcsec\ and $\mathrm{PA}
= 230\arcdeg$ a large shock feature can be detected in the ACS
f658n mosaic (labeled Southwest Shock in Figure~\ref{fig:bob3}),
which has a symmetry axis pointed towards $\mathrm{PA} =
220\arcdeg$. It is possible that this shock is associated with this
southwest SiO+CO outflow from \OW{137}{408}.

As mentioned in \S~3.6, HH 400 lies farthest from the cluster center
of all cataloged HH objects.  Figure~\ref{fig:bob3} illustrates that
HH~400 lies almost on a projection of the symmetry axes of three inner
HH objects (HH~203, HH~204, and HH~528). Since we know from the
curving of bipolar jets in this same part of the nebula \citep{bal06}
that there is a distortion of flows by a low velocity wind, it is
entirely possible that the small difference in alignment is due to
this effect. Unfortunately, this leaves identification of an
association with a specific inner HH object impossible.  The exact
location of HH 400 along the line of sight is very uncertain,
especially because of the report of \citet{bal01} that the shock is
redshifted with respect to the nearby nebular \SII{} emission.  Beyond
HH~400 are two groups of shocks, labeled 400-S and 400-SS in
Figure~\ref{fig:bob3}, and which are only seen in the ACS f658n
images. Due to their proximity and near alignment with HH 400, it
seems highly likely that they are produced by earlier ejections of
collimated material. This would mean that there is evidence for four
periods of intense outflow in this direction.

The ACS mosaic also reveals a large set of shocks that appear in
Figure~\ref{fig:bob3} and are labeled Western Shocks. The northernmost
members of this group could be related to the HH~269 outflow. Although
they cover a wide range of angles, this could be explained by episodic
ejection striking a deflecting surface that has changed between the
periods of outflow, hence one would expect both different radial
velocity and PA values. The much less well defined Northwest Shocks
features are unlikely to be related to HH~625, since it is seen to be
rotating clockwise with increased distance from its source; but, could
be related to either the outflow producing HH~202 or the fingers seen
around BN-KL.

\subsection{Origin of the Remaining HH Objects and Comparison with
  Earlier Studies}

In the above discussions we have examined possible origins of many of
the cataloged HH objects in the inner part of the Orion
Nebula. However, the origin of several major features remains
open. The opposite directions of the tangential velocities of the
innermost parts of HH 269 and HH 529 \citepalias{ode03} indicates that
if they share a common origin that this lies in the region identified
as the OOS. The fact that projections of the directions of the high
velocity jets leading to HH 202 and HH 203 pass through the OOS led
\citetalias{ode03} to speculate that it is a center of their outflows
too. Since the symmetry axis of HH 528 also crosses the OOS, it too
could originate there.  However, in this paper we have shown that it
can be argued, albeit not definitively, that HH 203, HH 204, and/or HH
528 could be a result of the deflection of the originally redshifted
SiO outflow from \OW{136}{356}. This means that we will limit the
discussion here to HH 202, HH 269, and HH 529.  In
Figure~\ref{fig:bob5} we see the structure of the jet feeding HH
202. Its location demonstrates that the source is not the star
\OW{138}{340} postulated by \citet{smi04} from their Fabry-Perot
velocity maps of this region. Moreover, their Figure 2 shows that the
HH 202 jet passes slightly east of \OW{138}{340}, which is clearer in
the HeI velocity picture shown in our Figure~\ref{fig:bob5} and in
\citetalias{doi04}'s Figure~16.  \citet{smi04} also argue on the basis
of the opposing alignment that HH~202 and HH~528 are part of the same
outflow, with HH~528 being the originally redshifted
component. Similar opposing alignments with a large intervening gap
seem to be an insufficient argument for establishing such a
connection. While a deflection off the PDR will largely cause a
shifting of the velocity along the line of sight, it is also expected
to produce some change in direction in the plane of the sky.  We note
that the southern part of HH~202's high velocity jet curves clockwise
at its southern end, indicating that it may not originate within the
OOS. However, since jets can curve more than once, the possibility
remains open.  \citet{zap04} argued that the elongation of the 1.3 cm
source \OW{143}{353} (which falls within the OOS) towards HH~202 means
that this is the source of HH~202's jet. However, for this to be the
case, the jet would have had to be bent several times to be displaced
and reoriented at the point where we begin to optically see it.

The alignment of the direction of the tangential velocities of the
westward moving objects ascribed by \citetalias{ode03} to HH~269
argues that they are part of the same flow. Since the more easterly
group lies immediately west of the OOS and east of the center of the
SiO outflow, they both cannot arise from the SiO outflow. However, the
more westerly HH~269 features are positions that are compatible with a
deflected SiO outflow. If that is the case, then the HH~269 components
lying even further to the west (beyond the region covered in
\citetalias{ode03}) could also have the same source. This
interpretation means dis-associating the two groups studied by
\citetalias{ode03} and arguing that the alignment of tangential
velocities and positions is only fortuitous. This may be the case, as
the east and west groups have quite different tangential velocities
($49\pm25~\kms$ and $16\pm5~\kms$ respectively. Definitive proof of
one interpretation or the other is lacking, but would be provided if
one could determine detect optical objects moving from the direction
of the SiO source and curing into the direction of the western
components of HH~269.

\citet{smi04} also argue that the projection of the high velocity
features associated with the western part of HH 529 intercepts the
infrared source \OW{144}{351} and that it is the source of the outflow
driving HH~529. This otherwise attractive argument is diminished by
the high spatial resolution HST tangential velocity study
\citepalias{ode03} which indicates that the common motion of the
features associated with the high velocity flow project about
2\arcsec\ south of \OW{144}{351}.

This means that what is happening in the OOS region and its
possible relationship to the major HH objects in the Orion Nebula
remains very much uncertain at this time. The absence of strong radio,
infrared, x-ray, or optical sources within the OOS remains a mystery,
if one or more outflows originates from there.  What is most likely to
help in understanding this region are tangential velocities over a
wide range of ionization states, with emphasis on the lowest
ionization ions since their emission will originate closest to the IF,
the PDR, and presumably the embedded source.

\acknowledgments

We are grateful to Takao Doi for sharing his KPNO \SII{} data prior to
publication, to Michael Richer for assistance during some of the SPM
observations, and to Jane Arthur and John Meaburn for helpful
discussions. Eddie Bergeron of the Space Telescope Science Institute was
particularly helpful in preparing the enormous volume of new Orion
ACS data.

Support for this study was provided in part by grant GO~10246 from the
Space Telescope Science Institute, and partly from DGAPA-UNAM, Mexico,
through projects IN115202 and IN112006, and CONACyT, Mexico, through a
research studentship to MTGD\@.  LFR and LAZ acknowledge the support
of DGAPA, UNAM, and of CONACyT (Mexico).

% {\it Facilities:} \facility{HST (ACS)}

\end{document}